\documentstyle[aps,twocolumn,epsf,floats]{revtex}

\begin{document}
\draft
\title
{Effective field theory for the small-$x$ evolution\\}
\author{I. Balitsky }
\address{ Physics Department, Old Dominion University, Norfolk 
VA 23529 \\
and Theory Group, Jefferson Lab, Newport News VA 23606}
\date{\today}
\maketitle

\begin{abstract}
The small-$x$ behavior of structure functions in the saturation 
region is determined by the non-linear generalization of the BFKL 
equation. I suggest the effective field theory for the small-$x$ 
evolution which solves formally this equation. The result is the 
$2+1$ functional integral for the structure functions at small $x$. 
\end{abstract}



\narrowtext
 The great success of pQCD in describing the $Q^2$ behavior of structure 
functions of deep inelastic scattering (DIS) can be traced back to the fact 
that the $Q^2$ dependence is
governed by DGLAP evolution equations which have two remarkable properties: 
they are linear equations, and the evolution at high $Q^2$ is purely
perturbative (the  non-perturbative physics enters the 
game only when we lower the normalization point $\mu^2$ down to the typical 
hadronic scale $\sim 1$GeV). The higher-order terms of perturbative expansion 
for both the coefficient functions and the
anomalous dimensions of the light-cone operators lie in 
the same framework of linear evolution and lead to the corrections
$\sim\alpha_s,\alpha_s^2$ etc. 

The situation for the small-x DIS is more complicated.
 In the leading logarithmic approximation (LLA)
the small-x
asymptotics  is described by the
BFKL pomeron\cite{bfkl}.
It is possible to reformulate the BFKL equation as an evolution equation where
the relevant operators are Wilson lines - infinite gauge
links\cite{np:b463:99}.  
\footnote{
At high energies the particles move so fast
that their  trajectories can
be approximated by straight lines collinear to their velocities. 
The proper degrees of freedom for the
fast particles moving along the straight lines are the (infinite)  
gauge factors  ordered along the straight line \cite{nacht}.}
 The evolution of
the two-Wilson-line operator (``color dipole'')  with respect to the slope 
of Wilson lines reproduces the BFKL equation. 

Unfortunately, the theoretical status of the BFKL evolution is not as clear as
the DGLAP one (for the review, see Refs.
\cite{obzors}). The biggest problem is the lack of unitarity: the power
behavior of the BFKL cross section violates the Froissart bound and 
therefore, in 
order get the true asymptotics at small $x$, we must go beyond the LLA. 
At this step, we face a new problem. 
In the DGLAP case,
the sub-leading logaritms follow the same general pattern of linear DGLAP
equation and the problem is purely technical: 
calculating the loop corrections to 
the kernels.  In the case of small-$x$ evolution there are also $\alpha_s$
corrections to the BFKL kernel\cite{nlobfkl}, but, on the top
of that, there are the unitarity corrections which lie outside the framework of
the BFKL equation. At small $\alpha_s$ and $x$, these corrections seem to 
dominate over the NLO BFKL 
ones\cite{pl:b439:428}.

Another problem with the BFKL evolution is infrared 
instability.  We can safely apply pQCD to the small-$x$ DIS if 
the characteristic transverse momenta of the gluons $k_{\perp}$
in the gluon ladder are large. For the the first few
diagrams, one can check by explicit calculation that the
characteristic $k^2_{\perp}$ are $\sim Q^2$. However, as $x$ decreases, it
turns out that the characteristic transverse momenta in the middle of the gluon
ladder drift to $\Lambda_{\rm QCD}$ making the application of pQCD
questionable. 

Recently, an idea has emerged that these two difficulties may cancel each
other out. Consider the DIS from the heavy nuclei where
the large density sets the saturation scale $Q_s$
\cite{prep:100:1,np:b268:427,detz,np:b558:285} which effectively cuts the
integration over $k_{\perp}$ even at relatively low energy. As we shall see
below, the small-$x$ evolution in this case is non-linear which leads to the 
growth of the saturation scale with energy, see the
discussion in Refs. 
\cite{prep:100:1,np:b268:427,detz,np:b558:285,yura,np:b573:833}. 
It is natural to 
assume that even for the DIS from the nucleon
where there is no saturation at low energies, the saturation scale at
sufficiently small $x$ may be  generated by the non-linear evolution itself.
Indeed,  the parton recombination described by
the non-linear evolution must balance at some point the effects of 
parton splitting so the
partons will reach the state of the saturation. 
In this high-density regime the
coupling constant is small but the characteristic fields are large,  making a
perfect case for the application of the semiclassical QCD
methods\cite{detz,jklw,fizrev}. The high-density
regime of QCD can serve as a bridge between the domain of pQCD and the ``real''
non-perturbative QCD regime governed by the physics of confinement.

In this paper I suggest the effective field theory which describes the
small-$x$ evolution in the saturation region.  First, let me remind 
the OPE for
high-energy amplitudes derived  in  \cite{np:b463:99}. Consider the amplitude of
forward  $\gamma^*\gamma^*$-scattering at small $x_B={Q^2\over s}$.
In the target frame, the virtual photon splits into $q\bar{q}$ pair
which approaches the nucleon at high speed. Due to the high 
speed  the classical trajectories of the quarks are straight lines
collinear to the momentum of the incoming photon $q$. 
The corresponding operator expansion 
switched between nucleon states has the form \cite {np:b463:99}:
\begin{eqnarray}
&&\int \! d^{4}x
    e^{iq\cdot x} \langle p|T\{j_{\mu }(x)j_{\nu }(0)\}|p\rangle\nonumber\\
&&= ~\int d^{2}x_{\perp}
   I_{\mu\nu}(x_{\perp})\langle p|\mathop{\rm Tr}\{\hat{U}(x_{\perp})
\hat{U}^{\dagger }(0)\} |p\rangle,
\label{1.1}
\end{eqnarray}
where $I_{\mu\nu}(x_{\perp})$ is a certain numerical 
function of the transverse separation of quarks $x_{\perp}$ and virtuality
of the photon $Q^2=-q^2$. The relevant
operators $U(U^{\dagger })$ are gauge factors ordered along the classical
trajectories which are almost light-like lines stretching from minus to 
plus infinity:
\begin{equation}
U(z_{\perp} )= P\exp\left(i\int_{-\infty}^{\infty}due^{\mu}A_{\mu}(ue+z_{\perp})\right)
 \label{1.2}
 \end{equation}
where $e$ is collinear to $q$ and $z_{\perp}$ is the transverse position
of the Wilson line. 

It
turns out that the small-$x$ behavior of structure functions
is governed by the evolution of these operators with respect
to the deviation of the Wilson lines from the light cone; this
deviation serves as a kind of ``renormalization point" for these operators.
At infinite energy, the vector $e$ is light-like and
the corresponding matrix elements of the operators (\ref{1.2}) have a
logarithmic divergence in longitudinal momenta. To regularize it , 
we consider
operators corresponding to large but finite velocity and take 
$e_{\zeta}=e_1~+~\zeta e_2$
where $e_1=(q-\frac{q^2}{2pq}p)$ and $e_2=p$ are the lightlike 
vectors close to the directions of the colliding particles. 
Now, instead of studying the energy-dependence of the 
physical amplitude we must investigate the dependence of the operators
(\ref{1.2}) on the slope $\zeta$. Large energies mean small 
$\zeta$ and we can
sum up logarithms of $\zeta$ instead of logarithms of $s$
(At present, we can do it only in the leading logarithmic 
approximation (LLA) $\alpha_s\ll 1$, $\alpha_s\ln{s\over m^2}\sim 1$).
The equation governing
the dependense of $U$ on $\zeta$ has the form \cite{np:b463:99}
\footnote
{The first non-linear equation for parton densities is known 
since 1983 as the GLR equation  
(it was conjectured in Ref. \cite{prep:100:1} and proved in the double-log 
limit in 
Ref.~\cite{np:b268:427}). The full LLA $x$ result was first derived in Ref.
\cite{np:b463:99} by the above method. After that, it was  
reobtained in Ref.~\cite{yura} in the framework of the dipole 
model\cite{mu94,nnn}, 
in Ref. \cite{epj:c16:337} by direct summation of relevant Feynman diagrams,
and in Refs. \cite{iancu,wei} by the semiclassical
methods.} 
\begin{eqnarray}
\lefteqn{\zeta \frac{d}{d\zeta} {\cal
U}(x_{\perp},y_{\perp})}\label{master}\\
&=&{\alpha_sN_c\over 2\pi^2} \int
dz_{\perp} {(x_{\perp} -y_{\perp} )^2\over (x_{\perp} -z_{\perp} )^{2}
(z_{\perp} -y_{\perp} )^2}
\Big\{{\cal U}(x_{\perp},z_{\perp})\nonumber \\
&+&{\cal U}(z_{\perp},y_{\perp})-{\cal U}(x_{\perp},y_{\perp})+
{\cal U}(x_{\perp},z_{\perp}){\cal U}(z_{\perp},y_{\perp})\Big\}
\nonumber 
\end{eqnarray}
where ${\cal U}(x_{\perp},y_{\perp})
\equiv{1\over N_c}\mathop{\rm Tr}\{U(x_{\perp} )U^{\dagger } (y_{\perp} )\}-1$.
The first three linear terms in braces in the r.h.s. of eq. (\ref{master}))
reproduces the BFKL pomeron\cite{bfkl} while the quadratic term will give us the 
three-pomeron vertex\cite{bar3pom}. 
The solution of the linearized evolution equation is especially simple in 
the case of zero momentum transfer  (e.g. for the total 
cross section of small-x DIS):
\begin{eqnarray}
&&\langle p|{\cal U}^{\zeta=x_B}(x_{\perp},0)|p\rangle~=~
\int \! \frac {d\nu }{2\pi ^{2}}
(x_{\perp}^{2})^{-\frac {1}{2}+i\nu }\label{1.10}\\
&&
\left(\frac {s}{m^{2}}\right)^{\omega(\nu )}
\int \! dz_{\perp}(z_{\perp}^{2})^{-\frac {1}{2}-i\nu }
\langle p|{\cal U}^{\zeta_0}(z_{\perp},0)|p\rangle
\nonumber
\end{eqnarray} 
where 
$\omega(\nu )=2N_c{\alpha _{s}\over\pi }[-\mathop{\rm Re}\psi 
(\frac {1}{2}+i\nu )-C]$
and
$m^2$ is either $Q^2$ or $m_N^2$ (in LLA, we cannot distinquish
between $\alpha_s\ln{s\over Q^2}$ and $\alpha_s\ln{s\over m_N^2}$).
The sketch of linear evolution is presented in Fig. 1.
\begin{figure}
\vspace{-0cm}
\centerline{\epsfxsize 8.0cm\epsffile{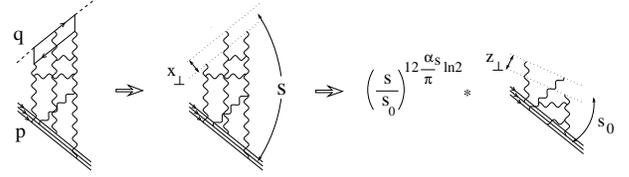}}
\vspace{5mm}
\caption[]{BFKL evolution in terms of Wilson-line operators
(denoted by dotted lines).
}
\end{figure}
The starting point of the evolution is the slope collinear 
to the momentum of the incoming photon q ($\zeta=x_B$)
and it is convenient to stop the evolution at a certain intermediate 
point $\zeta_0={Q^2\over s_0}$ where 
$s_0\gg m_N^2, ~~{\alpha_s\over\pi}\ln{s_0\over m_N^2}\ll 1~$.
The first of these conditions means that $s_0$ is still high from the 
viewpoint of low-energy
nucleon physics while the second condition means that $s_0$ is sufficiently
small from the viewpoint of high-energy physics 
(so one can neglect the BFKL logs).
The matrix element of the double-Wilson-line operator at this
slope is a phenomenological input for the BFKL evolution (just as the 
structure function at low $Q^2$ serves as the input for ordinary DGLAP evolution).
At large $s$ the integral over $\nu$ is dominated by the vicinity of $\nu=0$
which gives the familiar BFKL asymptotics
$\sigma^{\rm tot}\simeq x_B^{-12{\alpha_s\over \pi}\ln 2}$. 

Unlike the linear evolution, the general picture is very
complicated since the number of operators $U$ and $U^{\dagger }$ 
increases after each
evolution. At the time being, it is not known how to solve the non-linear
evolution equation in an explicit form. 
It is possible,
however, to write down the solution of the non-linear equation 
(\ref{master})
in the form of a functional integral over the double set of the
variables, 
$\varsigma_{i=1,2}(z_{\perp},\eta)=t^a\varsigma_{i=1,2}^a(z_{\perp},\eta)$ 
belonging
to the Lie
algebra of the SU(3) color group and $\Omega_{i=1,2}(z_{\perp},\eta)$ 
belonging to the
group itself: 
\begin{eqnarray}
&&\hspace{-0mm} U^{\eta_A}(x_{\perp})\otimes U^{\dagger\eta_A}(y_{\perp})~=
\label{yaugadal}\\
&&\hspace{-0mm} ~
\int^{\pi_{1,2}(\eta_A)=0}_{\Omega_{1,2}(\eta_0)=1}
D\varsigma_1(z,\eta) D\varsigma_2(z,\eta)
D\Omega_1(z,\eta) D\Omega_2(z,\eta)\nonumber\\
&&\hspace{-0mm} 
\Omega^{\dagger}_1(x_{\perp},\eta_A)
U^{\eta_0}_x\Omega_2(x_{\perp},\eta_A)\otimes 
\Omega^{\dagger}_2(y_{\perp},\eta_A)U^{\dagger\eta_0}_y
\Omega_1(y_{\perp},\eta_A)\nonumber\\
&&\hspace{-0mm} 
~\exp\Bigg\{\int^{\eta_A}_{\eta_0} d\eta \!\int\!d^2z
\Big[\frac{1}{g}\sum_{i=1,2}
\varsigma_i^a(z,\eta)\vec{\partial}^2
\Big(\Omega_i^{\dagger}(z,\eta)
\dot{\Omega}_i(z,\eta)\Big)^a\nonumber\\
&&\hspace{-0mm}~- 
\frac{1}{4\pi}\varsigma_1^a(z,\eta)\varsigma_2^b(z,\eta)
\vec{\partial}^2\Big(
\Omega_1^{\dagger}(z,\eta)U^{\eta_0}_z
\Omega_2(z,\eta)\Big)^{ab} \Big]\Bigg\}
 \nonumber 
\end{eqnarray}
where
$\dot{\Omega}\equiv\frac{\partial}{\partial\eta}\Omega$ and 
$\big(\Omega^{\dagger}\dot{\Omega}\big)^a\equiv  
2{\rm Tr}\{t^a\Omega^{\dagger}\dot{\Omega}\}$.  Going to
the  the variables $\pi=\vec{\partial}^2_{\perp}\varsigma$ we see that Eq.
(\ref{yaugadal}) is a phase-space  functional integral for the non-local
Hamiltonian 
\begin{eqnarray}
&&\hat{H}(\pi_1,\pi_2,\Omega_1, \Omega_2)=\label{ham}\\
&&
\int dx_{\perp}dy_{\perp}\pi_1^a(x_{\perp})
\hbox{\bf\big($\!\!$\big(} x_{\perp}
\big|\frac{1}{\vec{p}_{\perp}^2}
\big[\vec{\partial}_{\perp}^2\big(\Omega_1^{\dagger}
\Omega_2\big)^{ab}\big]\frac{1}{\vec{p}_{\perp}^2}
\big|y_{\perp}\hbox{\bf\big)$\!\!$\big)}
\pi_2^b(y_{\perp})
\nonumber
\end{eqnarray}
where $\big|x\hbox{\bf\big)$\!\!$\big)}$
is an eigenstate of the coordinate operator normalized according to 
$\hbox{\bf\big($\!\!$\big(} x
\big|y\hbox{\bf\big)$\!\!$\big)}=\delta^{(2)}(x-y)$, 
see e.g Ref. \cite{eveq}.
The rapidity $\eta$ serves as a Euclidean ``time'' for 
this evolution. 

We shall demonstrate that the perturbative expansion of the functional integral
 (\ref{yaugadal}) reproduces the evolution of the color dipole 
$U(x_{\perp})\otimes U^{\dagger}(y_{\perp})$ in the LLA. To get the
perturbative series, we substitute
$\Omega(x_{\perp},\eta)=e^{-ig\phi(x_{\perp},\eta)}$: 
\begin{eqnarray}
&&\hspace{-0mm} U^{\eta_A}_x\otimes U^{\dagger\eta_A}_y~=~
\int^{\pi_{1,2}(\eta_A)=0}_{\phi_{1,2}(\eta_0)=0}
\Pi_{i=1,2}D\varsigma_i(z,\eta) 
D\phi_i(z,\eta) \nonumber  \\
&&\hspace{-0mm}\times~ 
e^{ig\phi_1(x_{\perp},\eta_A)}
U^{\eta_0}_x e^{-ig\phi_2(x_{\perp},\eta_A)}\otimes 
e^{ig\phi_2(y_{\perp},\eta_A)}U^{\dagger\eta_0}_y
\nonumber\\ 
&&\hspace{-0mm}\times~e^{-ig\phi_1(y_{\perp},\eta_A)}~
\exp\Bigg\{\int^{\eta_A}_{\eta_0} d\eta \!\int\!d^2z
\Big[\frac{1}{g}\sum_{i=1,2}
\varsigma_i^a(z,\eta)\nonumber\\
&&~~~~~~~~~~~~~~~~~~~~~~\times~\vec{\partial}^2
\Big(e^{ig\phi_i(z,\eta)}\frac{\partial}{\partial\eta}
e^{-ig\phi_i(z,\eta)}\Big)^a
\nonumber\\ 
&&\hspace{-0mm} -~ 
\frac{1}{4\pi}\varsigma_1^a(z,\eta)\varsigma_2^b(z,\eta)
\vec{\partial}^2\Big(
e^{ig\phi_1(z,\eta)}U^{\eta_0}_z
e^{-ig\phi_2(z,\eta)}\Big)^{ab} \Big]\Bigg\}
\label{ugadalo}
\end{eqnarray}
Next, we can represent the
r.h.s. of Eq. (\ref{ugadalo}) in the 
form ($\dot{\phi}\equiv\frac{\partial\phi}{\partial\eta}$)
\begin{eqnarray}
&&\hspace{-0mm}
\int^{\pi_{1,2}(\eta_A)=0}_{\phi_{1,2}(\eta_0)=0}
\Pi_{i=1,2}D\varsigma_i(z,\eta) 
D\phi_i(z,\eta)\label{proverim1}\\
&&
[\eta_A,\eta_0]_xU^{\eta_0}_x\{\eta_0,\eta_A\}_x
\otimes~ 
\{\eta_A,\eta_0\}_y
U^{\dagger\eta_0}_y
[\eta_0,\eta_A]_y
\nonumber\\ 
&&\hspace{-0mm}\times~
\exp\Bigg\{\int^{\eta_A}_{\eta_0} d\eta \!\int\!d^2z
\Big[-i\sum_{i=1,2}
\varsigma_i^a(z,\eta)\vec{\partial}^2
\dot{\phi}_i(z,\eta)\nonumber\\ 
&&\hspace{-0mm}-~\frac{1}{4\pi}
\varsigma_1^a(z,\eta)\vec{\partial}^2
\Big([\eta_A,\eta_0]_zU^{\eta_0}_z\{\eta_0,\eta_A\}_z
\Big)^{ab}\varsigma_2^b(z,\eta)
\Big]\Bigg\}
\nonumber
\end{eqnarray}
where we introduced the notations
\begin{equation}
[\eta_1,\eta_2]_x\equiv
{\rm T}e^{ig\int^{\eta_1}_{\eta_2}
\dot{\phi}_1(x_{\perp},\eta)},~\{\eta_1,\eta_2\}_x\equiv
{\rm T}e^{ig\int^{\eta_1}_{\eta_2}
\dot{\phi}_2(x_{\perp},\eta)}
\label{notpexp}
\end{equation}
Let us now expand the r.h.s. of 
Eq. (\ref{proverim1}) in powers of $g$. The first
nontrivial term in this expansion is
\begin{eqnarray}
&&\hspace{-0mm} U^{\eta_A}_x\otimes U^{\dagger\eta_A}_y\label{firsterm}\\ 
&&\hspace{-0mm}=~
\alpha_s\int^{\pi_{1,2}(\eta_A)=0}_{\phi_{1,2}(\eta_0)=0}
\Pi_{i=1,2}D\varsigma_i(z,\eta) 
D\phi_i(z,\eta)\nonumber \\ 
&&\hspace{-0mm}\times~ 
\Bigg[\Big(\phi_1(x_{\perp},\eta_A)
U^{\eta_0}_x-U^{\eta_0}_x\phi_2(x_{\perp},\eta_A)\Big)\nonumber\\ 
&&\hspace{-0mm} 
\otimes~
\Big(\phi_2(y_{\perp},\eta_A)U^{\dagger\eta_0}_y-
U^{\dagger\eta_0}_y\phi_1(y_{\perp},\eta_A)\Big)
\nonumber\\ 
&&\hspace{-0mm}-~ 
\Big(\phi_1(x_{\perp},\eta_0)
U^{\eta_0}_x \phi_2(x_{\perp},\eta_A)\Big)
\otimes U^{\dagger\eta_0}_y\nonumber\\ 
&&\hspace{-0mm}+U^{\eta_0}_x\otimes
\Big(\phi_2(y_{\perp},\eta_A)U^{\dagger\eta_0}_y
\phi_1(y_{\perp},\eta_A)\Big)\Bigg]
\nonumber\\ 
&&\hspace{-0mm}\times\int^{\eta_A}_{\eta_0} d\eta \!\int\!d^2z
\varsigma_1^a(z,\eta)\varsigma_2^b(z,\eta)
\vec{\partial}^2\Big(
U^{\eta_0}(z_{\perp})\Big)^{ab} 
\nonumber\\
&&\times~
\exp\Bigg\{-i\int^{\eta_A}_{\eta_0} d\eta \!\int\!d^2z
\sum_{i=1,2}
\varsigma_i^a(z,\eta)\vec{\partial}^2
\dot{\phi}^a_i(z,\eta)
\Bigg\}.
\nonumber
\end{eqnarray}
The propagators for this phase-space functional integral are 
\begin{eqnarray}
&&\langle\phi^a_i(x,\eta)\varsigma^b_j(y,\eta')\rangle=
i\delta_{ij}\delta^{ab}
\hbox{\bf\Big($\!\!$\Big(} x
\Big|\frac{1}{\vec{p}^2}
\Big|y\hbox{\bf\Big)$\!\!$\Big)}\theta(\eta-\eta'),
\nonumber\\
&&\langle\phi^a_i(x,\eta)\phi_j^b(y,\eta')\rangle= 0,~~~
\langle\varsigma^a_i(x,\eta)\varsigma^b_j(y,\eta')\rangle=0
\label{fipiprop}
\end{eqnarray}

With these propagators, the r.h.s of Eq. (\ref{firsterm}) reduces to
\begin{eqnarray}
&&\hspace{-0mm}-\alpha_s(\eta_A-\eta_0)\Bigg[
\Big(t^aU^{\eta_0}_x\otimes t^bU^{\dagger\eta_0}_y+
U^{\eta_0}_xt^b\otimes U^{\dagger\eta_0}_yt^a\Big)\nonumber\\
&&\times~
\hbox{\bf\big($\!\!$\big(} x
\big|\frac{1}{\vec{p}^2}\vec{\partial}^2U^{\eta_0}
\frac{1}{\vec{p}^2}
\big|y\hbox{\bf\big)$\!\!$\big)}^{ab}-~
t^aU^{\eta_0}_xt^b\otimes U^{\dagger\eta_0}_y
\hbox{\bf\big($\!\!$\big(} x
\big|\frac{1}{\vec{p}^2}\big(\vec{\partial}^2U^{\eta_0}\big)\nonumber\\
&&\times~\frac{1}{\vec{p}^2}
\big|x\hbox{\bf\big)$\!\!$\big)}^{ab}-
U^{\eta_0}_x\otimes t^bU^{\dagger\eta_0}_yt^a
\hbox{\bf\big($\!\!$\big(} y
\big|\frac{1}{\vec{p}^2}\big(\vec{\partial}^2U^{\eta_0}\big)
\frac{1}{\vec{p}^2}
\big|y\hbox{\bf\big)$\!\!$\big)}^{ab}\Bigg]
\label{sovpal}
\end{eqnarray}
which coincides with the Eq.  (B17) from Ref. \cite{np:b463:99}. Taking
trace over the color dipole indices one reproduces the Eq. (\ref{master}).
Similarly, it can be demonstrated that further terms of
the expansion of eq. (\ref{proverim1}) in powers of 
$g$ repoduce the subsequent iterations
of the non-linear equation (\ref{master}). 

The intergral over $\pi$ variables can be 
easily performed resulting in:
\begin{eqnarray}
&&\hspace{-0mm} U^{\eta_A}(x_{\perp})\otimes
U^{\dagger\eta_A}(y_{\perp})\label{ugadalu}\\ 
&&\hspace{-0mm}=~ 
\int_{\Omega_{1,2}(\eta_0)=1}^{\dot{\Omega}_{1,2}(\eta_A)=0}
D\Omega_1(z,\eta) D\Omega_2(z,\eta)~
\Omega^{\dagger}_1(x_{\perp},\eta_A)\nonumber\\
&&\hspace{-0mm}\times~
U^{\eta_0}(x_{\perp})\Omega_2(x_{\perp},\eta_A)\otimes 
\Omega^{\dagger}_2(y_{\perp},\eta_A)U^{\dagger\eta_0}(y_{\perp})
\Omega_1(y_{\perp},\eta_A)\nonumber\\
&&\hspace{-0mm}\times~
\exp\Big\{-\frac{1}{\alpha_s}
\int^{\eta_A}_{\eta_0}\!\!\! d\eta \!\int\!d^2z 
\Big[\vec{\partial}^2
\Big(\Omega_2^{\dagger}(z,\eta)U^{\dagger\eta_0}_z
\Omega_1(z,\eta)\Big)\Big]^{-1}_{ab}
\nonumber\\
&&\hspace{-0mm}\times~
\vec{\partial}^2
\Big(i\Omega_1^{\dagger}(z,\eta)
\dot{\Omega}_1(z,\eta)\Big)^a
\vec{\partial}^2
\Big(i\Omega_2^{\dagger}(z,\eta)
\dot{\Omega}_2(z,\eta)\Big)^b
\Big\}
 \nonumber 
\end{eqnarray}
Note that the action of this effective field theory is local. This functional
integral for the small-$x$ evolution of the Wilson-line operators is the main 
result of the paper.

In the case of large nuclei it is possible to write initial conditions
for the small-x evolution using the McLerran-Venugolalan model. 
The nuclear matrix element of the two-Wilson-line operator (``color dipole'')
is given by the Glauber formula, \cite{kop,mu90,myra} see Fig. 2.
\begin{figure}
\vspace{-0cm}
\centerline{\epsfxsize 8.0cm\epsffile{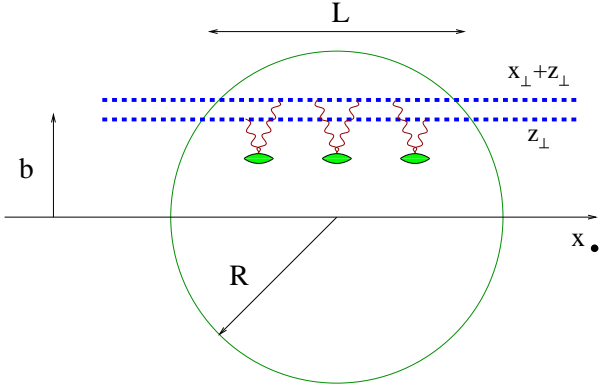}}
\vspace{0.5cm}
\caption[]{Propagation of the color dipole through the nucleus.
}
\end{figure}
\begin{eqnarray} 
&&\int d^2z_{\perp}\langle A|
{\rm Tr}U(x_{\perp}+z_{\perp})U(z_{\perp})|A\rangle\nonumber\\
&&=~
N_c\int
d^2b \left[1-e^{-g^2c_F{\cal G}(x_{\perp}^2)
L_b}\right]
\label{eq:bal:con2}
\end{eqnarray}
Here $L_b\equiv 2\sqrt{R^2-b^2}$ is the propagation
length of  the dipole (located at the impact parameter $b$) 
through the nucleus,
 $\rho=\frac{A}{4/3\pi R^3}$ is the
nuclear density, and  
\begin{equation} 
{\cal G}(x_{\perp}^2)\equiv \frac{\pi
x_{\perp}^2}{4(N_c^2-1)}\rho\sigma_0 
G(\sigma_0,\mu^2=\frac{1}{x_{\perp}^2}). 
\label{eq:bal:con3}
\end{equation}
The Eq. (\ref{eq:bal:con3}) is derived
under the assumption that the characteristic size of the dipole (the
``saturation scale '') is smaller than the size of the nucleon 
\footnote
{This assumption is certainly true at $A\rightarrow\infty$. For 
real nuclei, one should find the saturation scale $Q_s$ from the final result
for the matrix element of the color dipole between the nuclear states,
and verify that $Q_s\gg 1$GeV}.
In this case,  the quarks propagating along the
straight light-like lines
\footnote{As we mentioned above, the energy $s\sigma_0$ should
be high enough so we can replace the slope $p_1+\zeta_0 p_2$ by $p_1$ in the
non-logarithmical expressions.}
 interact by the instantaneous (in the light-cone
time $x_{+}$) potential 
\begin{eqnarray}
&&\rho g^2t^a\otimes t^a\int
\frac{d^2p_\perp}{(2\pi)^2}\frac{g^2}{2p_{\perp}^4}
\Big(e^{i(p,x-y)_{\perp}}-1\Big)\nonumber\\
&& =~ g^2t^a\otimes t^a\rho
\frac{\alpha_s}{8}(x-y)_{\perp}^2\ln(x-y)_{\perp}^2m_0^2
\label{eq:bal:con3a}
\end{eqnarray}
where $m_0\simeq\frac{m_N}{2}$ is the IR cutoff.\cite{mu90} It is worth noting that the factor $-1$ in the parenthesis in
the l.h.s. comes from the diagrams with the two gluons attached to the same
nucleon and the same Wilson line. Taking into account the color factors, one  obtains the Eq.
(\ref{eq:bal:con2}) with  $x_BG\big(x_B,\mu^2=x_{\perp}^{-2}\big)=
\frac{4\alpha_s}{\pi}\ln x_{\perp}^{-2}/m_0^2$, see Ref. \cite{mu90}. 

Similarly to Eq. (\ref{ugadalu}), it is possible to represent this result as a
functional integral  over a variable 
$\Lambda(x_{\perp},l)\in SU(3)$:
\begin{eqnarray}
&&\int d^2z\langle A|U^{\eta_0}_{x+z}U^{\dagger\eta_0}_{z}
|A\rangle
\label{con8}\\
&&=\int d^2b\int_{\Lambda(0,y)=1}^{\Lambda'(L_b,y)=0} 
 D\Lambda(y,l)~ \Lambda(x+z,L_b)~
\Lambda^{\dagger}(z,L_b)\nonumber\\
 &&\hspace{1cm}\times~
\exp\Big\{\frac{1}{2g^2\rho}\int_0^{L_b}\! dl\! \int\! d^2y
\big(\Lambda(l,y)
\Lambda'(l,y)\big)^a
\nonumber\\
&&\hspace{2cm}\times~
(-\vec{\partial}^2+m_0^2)^2
\big(\Lambda(l,y)
\Lambda'(l,y)\big)^a\Big\} \nonumber
\end{eqnarray}
where $\Lambda'\equiv{\partial\over\partial l}\Lambda$.
Extra $U^{\eta_0}(x)$  $\big(U^{\dagger\eta_0}(x)\big)$ lead to extra 
$\Lambda(x,L_b)$ $\big(\Lambda^{\dagger}(x,L_b)\big)$ in the pre-exponent.

The final formula for the martix element of the color dipole operator
at small $x_B$ is obtained by combining the functional integrals 
(\ref{ugadalu}) 
and (\ref{con8}):
\begin{eqnarray}
&&\hspace{-0mm} \int d^2z\langle A|U^{\eta_A}(x+z)
\otimes
U^{\dagger\eta_A}(z)|A\rangle\nonumber \\ 
&&\hspace{-0mm}=~\int\! d^2b\! 
\int_{\Lambda(0,y_{\perp})=1}^{\Lambda'(L_b,y_{\perp})=0} 
\int_{\Omega_{1,2}(\eta_0,y)=1}^{\dot{\Omega}_{1,2}(\eta_A,y)=0}
D\Lambda(l,y)D
\Omega_1(y,\eta)\nonumber\\
&&\hspace{-0mm}\times~
 D\Omega_2(y,\eta)~
\Omega^{\dagger}_1(x+z,\eta_A)
\Lambda(L_b,x+z)\nonumber\\
&&\hspace{-0mm}\otimes~\Omega_2(x+z,\eta_A)
\Omega^{\dagger}_2(z,\eta_A)\Lambda^{\dagger}(L_b,z)
\Omega_1(z,\eta_A)\nonumber\\
&&\hspace{-0mm}\times~
\exp\Bigg\{-\frac{1}{2g^2\rho}\int_0^{L_b}\!\! dl \int d^2y
\big(i\Lambda(l,y)\Lambda'(l,y)\big)^a\nonumber\\
&&\hspace{25mm}\times~ (m_0^2-\vec{\partial}^2)^2
\big(i\Lambda(l,y_{\perp})\Lambda'(l,y)\big)^a \nonumber\\
&&\hspace{-0mm}- ~\frac{1}{\alpha_s}
\int^{\eta_A}_{\eta_0} d\eta \!\int\!d^2y
\vec{\partial}^2
\Big(i\Omega_1^{\dagger}(y,\eta)
\dot{\Omega}_1(y,\eta)\Big)^a
\nonumber\\
&&\hspace{15mm}\times~
\Big[\vec{\partial}^2
\Big(\Omega_2^{\dagger}(y,\eta)\Lambda^{\dagger}(L_b,y)
\Omega_1(y,\eta)\Big)\Big]^{-1}_{ab}\nonumber\\
&&\hspace{25mm}\times~
\vec{\partial}^2
\Big(i\Omega_2^{\dagger}(y,\eta) 
\dot{\Omega}_2(y,\eta)\Big)^b
\Bigg\}.
\label{otvet}
\end{eqnarray}
The gluon structure function 
in the LLA is proportional to the matrix element of the dipole 
operator
$x_BG\big(x_B,\mu^2=x_{\perp}^{-2}\big)=-\frac{2\pi}{s}\langle A|
{\rm Tr}U_i^{\eta_A}(x_{\perp})U_i^{\eta_A}(0)|A\rangle$, so the
numerical calculation of the functional integral (\ref{otvet}) should give 
the nuclear structure functions at small $x$. This would be complementary to
the approximate solutions of Refs. 
\cite{yura,epj:c16:337,np:b573:833,iancu,wei,ltu} since it could give the 
structure functions not only in the asymptotic black-body limit, but also in 
the intermediate region defining the saturation scale $Q_s$.
 
It should be mentioned that our formula (\ref{ugadalu}) 
gives the evolution of the color dipole only in the LLA. 
In the case of large nucleus we have an additional parameter $A\gg 1$ so our
LLA approximation based on the non-linear equation (\ref{master}) 
has a window 
$\alpha_s^2A^{1/3}\sim 1$, $\alpha_s\ln x_B\sim 1$ where it is justified even
at moderately small $x_B$. In the case of nucleon, our $\alpha_s(Q_s)\ll
1$,  $\alpha_s(Q_s)\ln x_B\sim 1$ approximation should be justified {\it a
posteriori}  after checking that the saturation does occur at sufficiently small
$x_B$. If the  saturation takes place at such low $x$ that $\alpha_s(Q_s)\ln
x_B\gg 1$, our LLA breaks down and we need to take into account the non-fan
diagrams such as t-channel loops formed by BFKL pomerons. However, 
the non-linear equation (\ref{master}) leads to the result for the
structure function which does not violate unitarity (see  the discussion in
Refs. \cite{yura,np:b573:833,jklw,epj:c16:337,np:b437:107})
and therefore we should not expect the large discrepancy between the unitary
LLA result and the exact amplitude at present energies.

\vskip0.5cm
\paragraph*{Acknowledgements.}
The author is grateful to Y.V. Kovchegov, E.M. Levin and L. McLerran 
for valuable discussions. 
This work was supported by the US Department of Energy under contract 
DE-AC05-84ER40150.

\end{document}